\documentclass{article}

\usepackage[english]{babel}

\usepackage[letterpaper,top=2cm,bottom=2cm,left=3cm,right=3cm,marginparwidth=1.75cm]{geometry}

\usepackage{amsmath}
\usepackage{graphicx}
\usepackage[colorlinks=true, allcolors=blue]{hyperref}
\usepackage{soul}
\usepackage{framed}

\providecommand{\keywords}[1]
{
  \small	
  \textbf{\textit{Keywords---}} #1
}

\title{Mental causation in a physical world: A self-causation model of downward causation}
\author{Yoshiyuki Ohmura and Yasuo Kuniyoshi}

\begin{document}
\maketitle

\begin{abstract}
Downward causation is self-causation, the causal effect from the whole to its parts, and is considered a promising theory for the problem of mental causation. 
However, it remains to be clarified how an irreducible but supervenient downward causal power can arise. 
Here, we argue that a feedback control of lower micro-level synaptic weights using higher macro-level algebraic structural feedback errors is a  model of downward causation. 
The feedback control consists of two mechanisms: observation of the feedback error and control of the feedback error. 
The two mechanisms can be implemented at two different levels of the  hierarchy. 
Macro-level feedback error is described by algebra of supervenient macro-level functions, independent of the external cause. 
The supervenient macro-level feedback error  constraining of the micro synaptic weights through feedback control is downward causation. 
\end{abstract}

\keywords{mind--brain problem, downward causation, algebraic structure control}

\section{Introduction}
\paragraph{Self-causation is necessary to explain mental causation} \quad \\
The mental causation problem in the mind--body problem asks how the mind can affect the body or brain. Mental causation is essential for explaining mental phenomena, including conscious experience, and free will, to avoid epiphenomenalism \cite{Baumeister2011,Baumeister2018}. However, the concept of mental causation suffers from a trilemma between the following three arguments\footnote{ We do not define the mind metaphysically, but the trilemma is described as independent of the definition of the mind.}.
\begin{itemize}
\item A1 (Physical causation): In physics, the causal object (the cause) and the affected object (the effect) are different physical entities.
\item A2 (Mental causation): The mind has a causal effect on the body or the brain. 
\item A3 (Mind-brain relation): The mind cannot be considered a separate physical entity distinct from the brain, because we have no empirical evidence for the physical mind. 
\end{itemize}
Descartes argued that the nature of the mind (i.e., a thinking, non-extended thing) is completely different from the nature of the body (i.e., an extended, non-thinking thing), where the extended thing occupies the space. For those who advocate for mental causation, Descartes' substance dualism is unacceptable. If the mind is not physical, we cannot explain causality from mind to body. Despite this, Descartes' concept does acknowledged the lack of evidence for a physical entity corresponding to the mind, such as a `quantum of mind'. 

We can observe when the brain is disturbed, then the mind changes. But we are observing inside the brain, we cannot observe the physical mind. This means that if the mind is generated or emerged from the brain and the mind can affect the brain, then the cause (the mind) and the effect (the brain) must share the same physical entities.

In a well-known exchange, Descartes' student, Princess Elisabeth pointed out that causality requires something to push and something to be pushed, and the pushed object and the pushing object must be different entities. A1 corresponds to Elisabeth's point.   
At present, many philosophers and scientists do not support substance dualism, but they do support A3 because we have no empirical evidence that the mind is a separate object from the brain. 
If we accept A3, then we must reject A1 in order to accept mental causation. 
To reject A1, we need to clarify the mechanism of self-causation, where we define self-causation as causation in which the cause and the effect share the same physical entity. Importantly, since our trilemma is not related to the completeness of physical causality, the existence of mental causation does not imply the incompleteness of physical causality. 

When the boundary of the physical system is clearly defined, the concept of self-causation has the same meaning as ``intrinsic'' cause, 
because intrinsic cause can affect the outside of the system by changing the system itself. 
For instance, intrinsic rewards -- such as an organism's internally driven motivations \cite{Cleeremans2022} -- can affect the surrounding environment after a change in behavior. 
In contrast, a non-living physical system cannot move without an external cause. 
Therefore, we believe that self-causation is one of the defining features of mental phenomena, including volition and conscious experience as suggested in \cite{Cleeremans2022}.
Regardless of its importance, no theoretical formulation of self-causation has been proposed. 
One promising approach to unravelling the mechanism of self-causation is the concept of downward causation proposed by Sperry \cite{sperry1964}. The concept seems to exemplify self-causation, which we call structural downward causation. 

\paragraph{Structural downward causation is self-causation} \quad \\
Structural downward causation is the causal effect from the whole to its parts \cite{popper1977, sperry1987, sperry1991, Oconer1994}. 
The brain is composed of a vast number of neurons, yet the mind is integrated, as we can see from our conscious experience \cite{tononi1998}.
If we assume that the brain is an interconnected causal network of micro-components, the boundaries of the causal network are not clear because the world outside the brain is also connected to the causal network.
This problem of boundary determination is called the problem of many \cite{Simon2018}. 
Importantly, biological systems are highly structured and have clear boundaries.  
To define boundary of self, we assume a hierarchical modular structure in the biological system at the physical level (i.e., embodiment). 
Modular structures are common in the hierarchical structure of the organism (i.e., the molecule, cell, tissue, organ, organism, population, and species). 
Since the whole organ (e.g., the  brain) and its component parts (e.g., neurons) share the same physical entities in the hierarchical structure, the causal effect from the more integrated macro modules to their components parts in the hierarchy is a plausible explanation for self-causation. 

Traditional micro-determinism reduces all biological phenomena to physics and chemistry, and this reductionism assumes that all macro-level facts and laws can be deduced from micro-level laws and initial conditions. 
From the reductionist view, structural downward causation is not possible, because the higher macro level is completely determined at the lower micro level. 
Reductionists assume that conscious experience lacks causal power and is epiphenomenal. 
In contrast to the traditional view, structural downward causation accepts that the whole system, which is physically composed of parts, has primordial causal power over its parts. 

\paragraph{The mechanism of structural downward causation is unclear} \quad \\
McLaughlin \cite{mclaughlin1992} argued that downward causation from either the biological or chemical level, as claimed by British emergentists and Sperry, is incorrect. Chalmers \cite{chalmers2006} considered that ``structural'' downward causation was theoretically reasonable but that no examples of ``structural'' downward causation are known in the actual world. 
Ellis gave the various examples of downward causation, but 
for each given case, the examples are not structural downward causation, because the higher and the lower levels of the example hierarchies do not share the same physical entity \cite{ellis2009, ellis2012,ellis2019}. 

In downward causation, it remains to be clarified how an irreducible but micro-dependent (supervenient) downward causal power arises \cite{Bedau1997}. Although the macro is determined by the micro, how can the macro have an autonomous causal effect on the micro? To address this question, we assume that there is feedback control at the micro level. To our knowledge, no one has ever examined whether downward causation is possible through feedback control.



\paragraph{Negative feedback control between different hierarchical levels} \quad \\ 
We assume that micro-level neurons are systems that can control their synapses through negative feedback control. 
Negative feedback control consists of two mechanisms: observation of the feedback error and control of the feedback error. Importantly, the two mechanisms can be implemented at two different levels of the hierarchy of the organism. 
We propose that a higher macro-level feedback error, which cannot be described at the lower micro-level alone (i.e., irreducibility), can  cause changes in the micro-level mechanism through feedback control. 

It is worth noting that, our formulation of structural downward causation is not related to self-organization and emergence in dynamical systems or non-linear complex systems \cite{Bedau1997,Gershenson2020}. 
In non-linear complex systems, the state-to-state transition under the fixed micro mechanism is often the focus and feedback control is reduced to attractors. Without distinguishing between feedback control and attractors, it becomes impossible to formulate structural downward causation, since our concept relies on recognizing that observation and control occur at different hierarchical levels. 
We provide a model of structural downward causation without emergence. 

The goal of typical feedback control is to converge a signal derived from measurement of  the system output to a reference signal (equal to the desired output). The feedback error is the difference between the reference and the actual output signal. Given that this reference is often provided from outside the system, the reference in the typical case is an external cause that does not depend on the micro. Such external causes are not suitable to define downward causation, because the macro-level feedback error must supervene on the micro\footnote{In this article, ``supervenience'' has the same meaning as micro-determination and micro-dependence.}. 

To define supervenient macro-level feedback error, we assume that the micro components are neurons and synaptic weights, and the macro components are mathematical functions (i.e., neural networks), not coarse-grained variables. 
We assume that a mathematical property supervenes on physical neurons and synapses, and algebra is used to define the supervenient macro-level feedback error. 

In this article, we propose that feedback control 
of the micro-level synaptic weights to satisfy the macro-level algebraic equation is structural downward causation.
Additionally we propose a supervenient macro-level feedback error that is irreducible to the micro-level, but plausible in brain-like systems. 

\section{Preliminaries}
Our structural downward causation formulation is based on three assumptions:
modular structure in the brain,
negative feedback control at the micro-level,
and macro-level feedback error based on algebraic equation. 

\paragraph{Modular structure in the brain} \quad \\
Bertalanffy \cite{bertalanffy1969} distinguished two biological hierarchies: structural and functional. An example of a biological structural hierarchy is the organization of the molecule, cell, tissue, organ, organism, population, and species. Meanwhile, a functional hierarchy is exemplified by the organization of the primary visual cortex, the secondary visual cortex, the association cortex, and the prefrontal cortex in the brain. 

The biological structural hierarchy relates the whole to its parts. Since the whole is physically composed of its micro-level parts, no additional material is required. Such a structural hierarchy defines structural downward causation. How the higher level acquires an irreducible but supervenient downward causal power is an open question, since this situation seems to be inconsistent with reductionism. 

A functional hierarchy, on the other hand, assumes different physical variables, systems or agents at both the higher and lower levels, so that causation from the higher to the lower level is not self-causation. In such cases, we refer to the non-supervenient macro as an external macro. 
Examples of the higher level system controlling the lower level system in the functional hierarchies are commonplace: a computer program in read-only memory controls a processor and memory outside the memory for the program, the central nervous system controls the peripheral nervous system, the motor cortex controls the muscle, and so on. 
To distinguish causation in functional hierarchies from structural downward causation, causation from the external macro to the micro in the functional hierarchy is here called functional downward causation \cite{ellis2012, ellis2019}. Ellis assumes that the higher variable in the hierarchy is a quantity that characterizes the state of the system \cite{ellis2009}. This indicates that Ellis' downward (top-down) causation is defined in a functional hierarchy. 
In the structural hierarchy, the higher level is not an abstract variable, but shares the same physical entities with lower level.
The way to distinguish structural downward causation from functional downward causation is to note whether the physical entity is shared by higher and lower levels in the hierarchy (e.g., motor cortex and muscle are different physical entities in the functional hierarchy). 

\begin{figure}[t]
\centering 
\includegraphics[width=1.\linewidth]{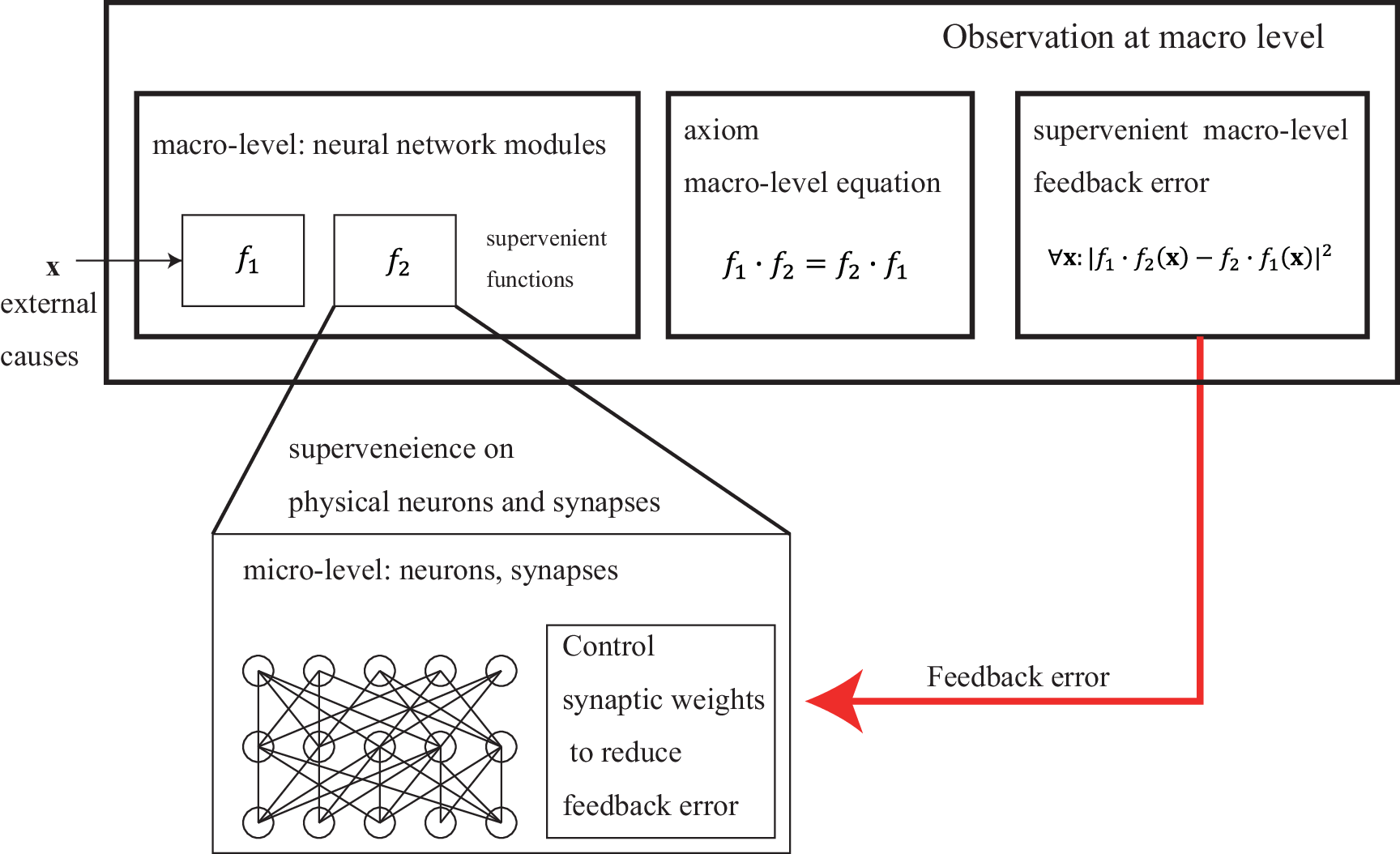}
\caption{\label{fig:model}
Structural downward causation model: At the macro level, there are several neural network modules. Each neural network module consists of a set of neurons and synapses at the micro level. And the macro functions supervenes on the micro.  
An axiom is a macro-level equation used to calculate the supervenient feedback error. The axiom is freely definable at the macro level due to multiple realizability. 
At the micro level, synaptic weights within each neural network module are controlled to reduce the feedback error.
Because the feedback error is described by supervenient functions independent of the external macro, we consider this mechanism as structural downward causation.}
\end{figure}

To formulate structural downward causation, we assume a structural hierarchy in the brain. The brain is physically composed of neurons and several types of glial cells. For simplicity, we will focus only on neurons.
A higher level organization must be physically composed of neurons and their synapses without any additional material to share the physical entity between higher and lower levels in the hierarchy.
Here, we assume the modular structure of neural networks at the macro level. Modular structures are most commonly found in the structural hierarchy of living organisms, and the neural network modules at the macro level are composed of a set of neurons and synapses (Figure \ref{fig:model}). 
The macro-level consists of a group of neural network modules, and the micro-level consists of the neurons and synapses within all the neural network modules, where the whole and its parts share the same physical entities. 
The columnar structures \cite{Mountcastle1997}, or memory engrams \cite{Josselyn2020}, in the cerebral cortex suggests that there are multiple neural network modules in the brain. 

\paragraph{Negative feedback control at the micro-level} \quad \\
Kim \cite{kim2006} argued that there is no causal power at the macro level because macro-level causality implies “overdetermination,” suggesting that the mind is an epiphenomenon; therefore, the whole structure cannot exert new causal power over the micro parts. 
This “causal exclusion” argument is often applied to argue against the possibility of mental causation beyond physical causation \cite{kim2000}, but it can be applied to all cases of supervenience, including the hierarchy of sciences \cite{bontly2002}. 
According to Bontly \cite{bontly2002}, Kim's position actually implies that only the properties of fundamental physical particles at the micro level are causally effective. 
To support structural downward causation, we must reject Kim's argument. 

Kim \cite{kim2006} assumed that higher level properties (corresponding to the macro) must directly cause the lower base properties (corresponding to the micro), without feedback control.  
However, we consider that the micro-level neurons actively control synaptic weights to reduce supervenient macro-level feedback error, rather than being passively controlled or superseded by the supervenient macro. 
This mechanism avoids ``overdetermination''. 
Without considering negative feedback control at the micro level, Kim's causal exclusion argument seems to be correct. 

We assume that the micro-level consists of synaptic weights, and the macro-level consists of mathematical functions (i.e., neural networks) that supervene on the physical neurons and synaptic weights. 
An equation to define feedback error is described by the supervenient functions at the macro-level, and this equation cannot be described at the micro-level alone. 
The equation to determine the supervenient feedback error can be freely determined at the macro-level but the crucial principle by which the equation is autonomously determined is not addressed in this article. 

\paragraph{Macro-level feedback error based on algebraic equation of the supervenient functions} \quad \\
We assume that the brain has multiple neural network modules at the macro level. 
Each neural network module is a mathematical function or map with plasticity.
A mathematical property of the functions supervenes on the synaptic weights. 

Note that a supervenient function ``alone'' cannot regulate the synaptic weights that make up the function itself due to supervenience. 
Macro level supervenient feedback error cannot be defined by any single neural network module because only the identity equation $f=f$ can be defined by the function $f$ and $=$ alone. 
Therefore, we assume a macro-level equation is described by multiple supervenient functions and algebraic operations (Figure \ref{fig:model}). 
We assume that the algebraic operations are static and not causal difference makers. 

An algebraic structure is defined by a set $S$ with operation ``$\cdot$'' in mathematics. 
We assume that a set is composed of the supervenient functions labeled by lowercase letters  \{ $f_1$, $f_2$, $f_3$ ...  \} . A binary operation combines any two elements $f_1$ and $f_2$ of $S$ to form an element of $S$, denoted $f_1 \cdot f_2$. 
We assume that a binary operation ``$\cdot$'' is a function composition between supervenient functions. 

An algebraic structure satisfies several requirements, known as axioms. The axioms are usually described in terms of  equations in group theory.  
To define such axioms, the function composition operation is required between the neural network modules. To satisfy this condition, the neural network modules must be transformations whose inputs and outputs are elements of a set such as vectors with the same dimension.
We assume that an input and output for the neural networks are elements of a set and labeled by bold letters \{ $\mathbf{x}$, $\mathbf{y}$ ... \}. 
Here, the input and output do not supervene on the synaptic weights.

The axioms are equations defined by a set of supervenient functions and the binary operations, and are independent of the input and output that does not supervene on the micro. 
The axioms can only be defined at the macro level because the modules are defined at the macro level structure. 
The algebraic structure cannot be defined at the micro level, indicating irreducibility. 
Furthermore,  the axioms to define algebraic structure can be described by the supervenient functions alone at the macro level, since the axioms are satisfied in all inputs and outputs. 

In this article, we argue that the brain system can evaluate an algebraic structural feedback error described by the supervenient functions at the macro level to regulate synaptic weights at micro level. 

\section{A formulation of structural downward causation in the brain}
In this section, we propose that an equation defining the algebraic structure between neural network modules can be used to measure supervenient feedback error at the macro level. 
We assume that the neural network module is mathematically modeled as a function.
And algebraic axioms is described by the neural network modules and algebraic operation. 
The algebraic structural feedback error to satisfy the algebraic axioms can be used for feedback control from the macro level to the micro level. 

In mathematics, axioms are prerequisites, but  we think of axioms as control targets that constrain the algebraic structure of the supervenient functions through feedback control. 

The algebraic structure of the supervenient functions, without feedback control, is not limited to the mathematical structures we commonly know. 
For example, commutativity is not satisfied by all algebraic structures. 
Commutativity is satisfied by products of integers and real numbers, but not by matrices.  
If commutativity holds between neural networks, it is because the brain has developed to satisfy the axiom through feedback control. 

In the brain, the axioms between neural network modules that are the target for feedback control can only be defined at the macro level. 
Therefore, a feedback error can only be evaluated at the macro level. 
Regulating its synaptic weights to reduce the feedback error is all the component neuron mechanisms do at micro level. 
The master controller to determine the target axioms and feedback error is derived at the macro level. This mechanism enables macro-determination. 
We call this mechanism algebraic structure control. 

In machine learning, these synaptic weights changes are enabled by gradient descent such as error back propagation \cite{rumelhart1986}. 
For example, consider commutativity axiom as a feedback error to satisfy a target algebraic structure and assume that $\mathbf{x}$ is input to two neural network modules, $f_1$ and $f_2$. 
The supervenient feedback error is defined as  $\forall \mathbf{x}:(f_1\cdot f_2(\mathbf{x})-f_2\cdot f_1(\mathbf{x}))^2$  \cite{ohmura2023}. 
To satisfy commutativity (which is not generally satisfied), the synaptic weights must be adjusted at the micro level. In this situation, the supervenient feedback error about the commutativity between neural network modules, $f_1$ and $f_2$, can be used to regulate the synaptic weights within the modules themselves at the micro level. 
This structural downward causation mechanism is logically plausible in the brain. 

\begin{figure}
\centering 
\includegraphics[width=1.\linewidth]{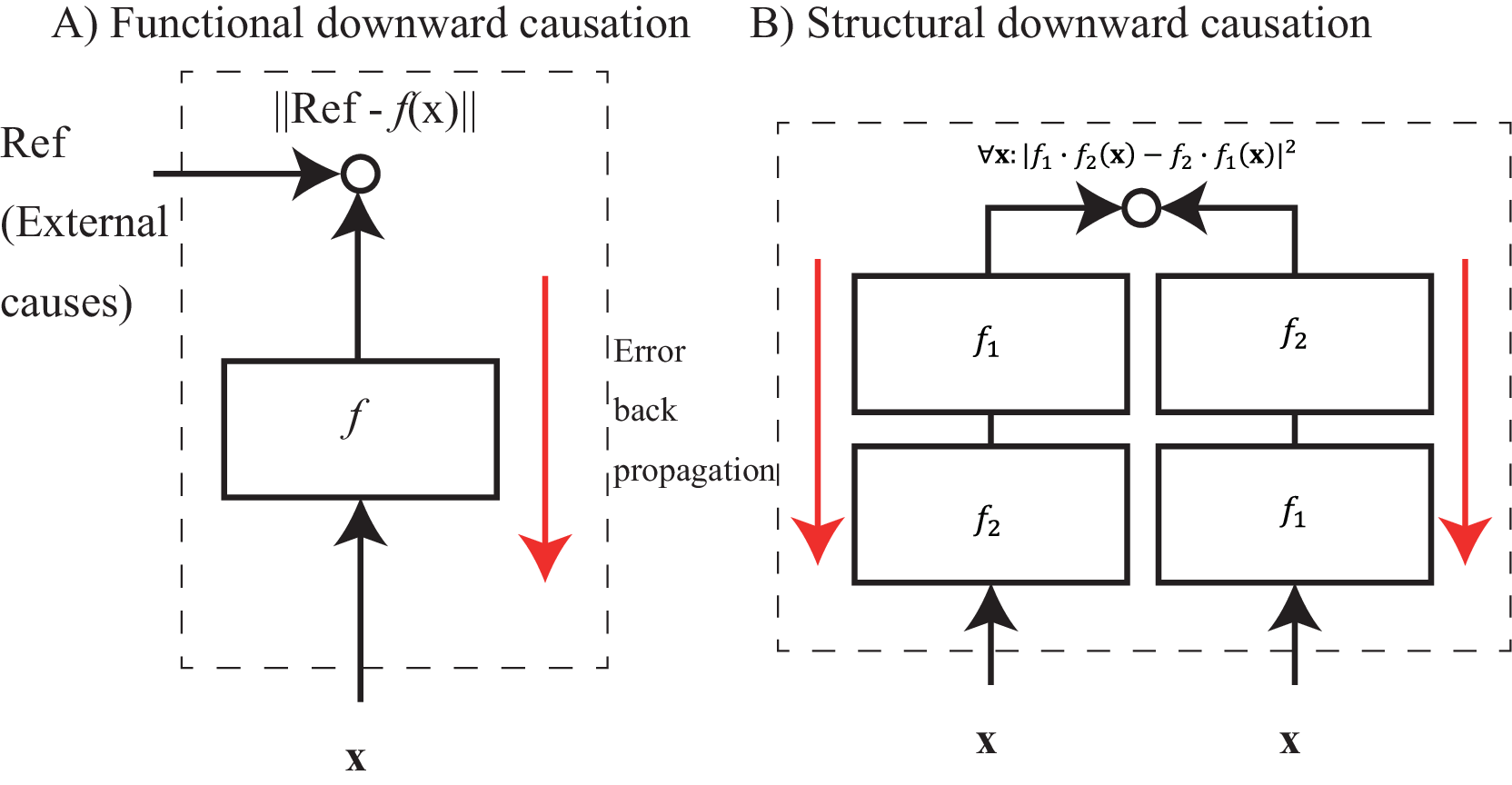}
\caption{\label{fig:outline}  The difference between functional downward causation and structural downward causation. 
Both update micro-level synaptic weights using macro-level description. 
However, defining supervenient feedback error using an external reference is problematic, since the relationship between the reference and the micro is not clear. 
(A) In functional downward causation, the feedback error is defined using a reference signal from the outside of the system. 
(B) In structural downward causation, the feedback error is defined by axioms to satisfy a target algebraic structure such as commutativity between the supervenient functions, $f_1$ and $f_2$}.   
\end{figure}

In conventional deep learning, a type of functional downward causation, the feedback error to control the synaptic weights depend on an external macro cause, including a reference signal supplied from the outside of the system (Figure \ref{fig:outline} A). 
In this case, the non-supervenient external macro is the cause of the synaptic weight changes at the micro level.  
In supervised learning, the labels of the data are given as a reference. 
In predictive learning, the data at different times are used as a reference.
In unsupervised learning, this is often done to approximate the data distribution to a probability distribution, such as a Gaussian distribution. 
The target and X do not supervene on the micro.
Therefore, the error between X and target is not a supervenient feedback error.

A unique feature of structural downward causation is that the feedback error is defined by the supervenient functions independent of an external macro (Figure \ref{fig:outline} B).
This feature makes it possible to generate supervenient feedback error at the macro level. 
Additionally the feedback error  can be evaluated inside of the system without need of reference signals from outside.

\section{Proof of concept}

\begin{figure}
\centering 
\includegraphics[width=1.\linewidth]{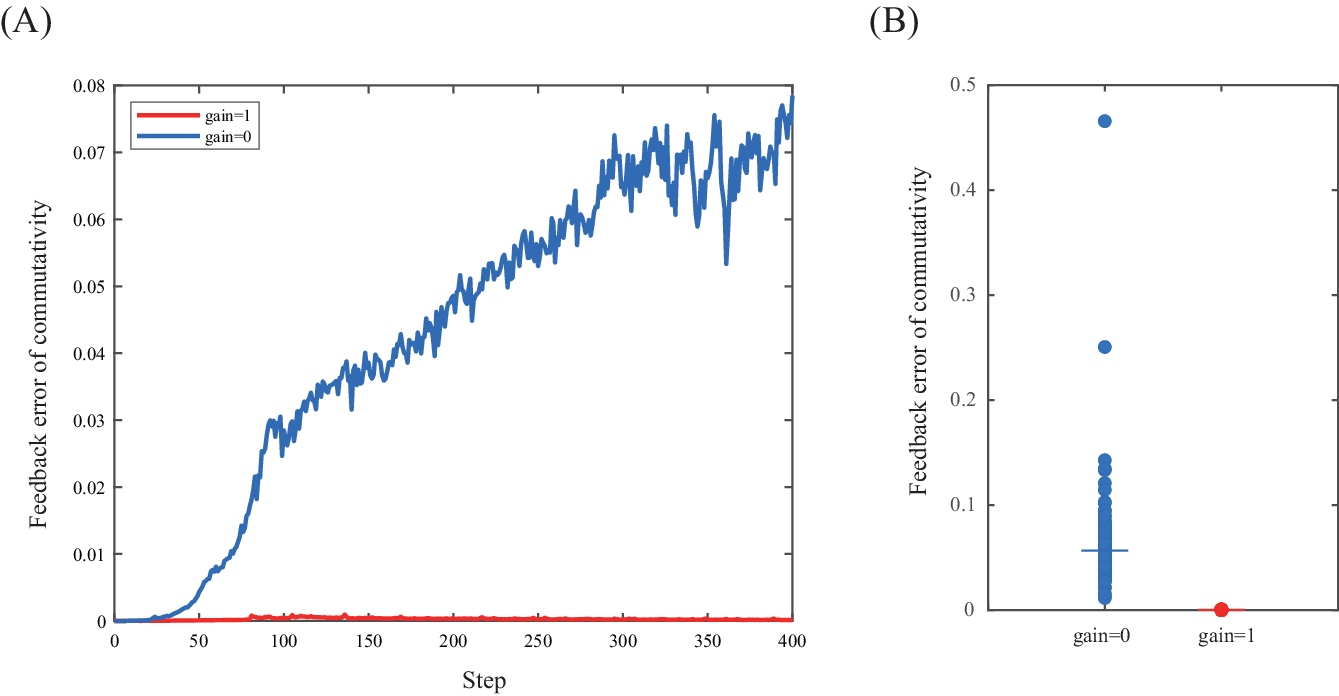}
\label{fig:result}  
\caption{
Feedback error of commutativity loss ($loss_c$).  
(A) When feedback gain is 0, the feedback error increased. In contrast, when feedback gain is 1, the feedback error converged to 0. Horizontal axis shows time step. 
(B)  Feedback error of commutativity is significantly different between feedback control condition ($gain$=1) and ablation condition ($gain$=0) (Mann-Whitney U = 10000, n=100, p $<$ 2.5e-4, two-tailed) The median of feedback error in feedback control condition is 1.76e-4. The median of feedback error in ablation condition is 0.056. 
}  
\end{figure}

We implement an example of structural downward causation. 
We used error back propagation \cite{rumelhart1986} to control synaptic weights. 
We show that commutativity at the macro level can be controlled by the synaptic weights at the micro level. 

We define a neural network that learns to convert input $\mathbf{x}$ to output $\mathbf{y}$. 
Without a supervenient feedback error at the macro level, the loss function is described by $\| \mathbf{y} - f[\boldsymbol{\lambda}](\mathbf{x}) \|$. 
To realize commutativity at the macro level, additional supervenient commutative loss, $\| f_1[\boldsymbol{\lambda}_1] \cdot f_2[\boldsymbol{\lambda}_2](\mathbf{x}) - f_2[\boldsymbol{\lambda}_2] \cdot f_1[\boldsymbol{\lambda}_1](\mathbf{x}) \| $ is required. 
Here, $\boldsymbol{\lambda}$, $\boldsymbol{\lambda}_1$, $\boldsymbol{\lambda}_2$ are vectors determined from $\mathbf{x}$ and $\mathbf{y}$.

We assume that the function $f[\boldsymbol{\lambda}]$ transforms from the whole input features  $\mathbf{x}$ to  the output features $\mathbf{y}$ for the conversion. 
The functions $f_1[\boldsymbol{\lambda}_1]$, $f_2[\boldsymbol{\lambda}_2]$ replace a part of the features and a combination of both $f_1$ and $f_2$ (i.e. $f_1[\boldsymbol{\lambda}_1] \cdot f_2[\boldsymbol{\lambda}_2]$ and $f_2[\boldsymbol{\lambda}_2] \cdot f_1[\boldsymbol{\lambda}_1]$) replace all the features (Figure \ref{fig:outline} b). 
To realize this, $\mathbf{x}$ is encoded into two latent vectors $x_1$ and $x_2$ (equation \ref{enc_x}). 
$\mathbf{y}$ is also encoded into two latent vectors $y_1$ and $y_2$ (equation \ref{enc_y}).

\begin{minipage}{0.9\hsize}
\begin{gather}
\label{enc_x}
x_1, x_2 = G_e (\mathbf{x}) \\
\label{enc_y}
y_1, y_2 = G_e (\mathbf{y}) 
\end{gather}
\end{minipage}

The function $f[\boldsymbol{\lambda}]$ replaces both $x_1$ and $x_2$ to $y_1$ and $y_2$. 
On the other hand, $f_1[\boldsymbol{\lambda}_1]$ replaces only $x_1$ to $y_1$, where $\boldsymbol{\lambda}_1 = y_1 - x_1$. 
Similarly, $f_2[\boldsymbol{\lambda}_2]$ replaces only $x_2$ to $y_2$, where $\boldsymbol{\lambda}_2 = y_2 - x_2$. 
The output $\mathbf{y}$, 
$f_1[\boldsymbol{\lambda}_1](\mathbf{x})$, and $f_2[\boldsymbol{\lambda}_2](\mathbf{x})$ 
are decoded from the latent vectors (equation
\ref{dec_y}, \ref{dec_f1x}, \ref{dec_f2x}). 

\begin{minipage}{0.9\hsize}
\begin{gather}
\label{dec_y}
\mathbf{y} = f[\boldsymbol{\lambda}](\mathbf{x}) = G_d (y_1, y_2) \\
\label{dec_f1x}
f_1[\boldsymbol{\lambda}_1](\mathbf{x}) = G_d (y_1, x_2) \\
\label{dec_f2x}
f_2[\boldsymbol{\lambda}_2](\mathbf{x}) = G_d (x_1, y_2)
\end{gather}
\end{minipage}

$f_1[\boldsymbol{\lambda}_1] \cdot f_2[\boldsymbol{\lambda}_2](\mathbf{x})$ and $f_2[\boldsymbol{\lambda}_2] \cdot f_1[\boldsymbol{\lambda}_1] (\mathbf{x})$ can be obtained using the same encoder and decoder. 

\begin{minipage}{0.9\hsize}
\begin{gather}
y_1^{'}, x_2^{'} = G_e (f_1[\boldsymbol{\lambda}_1](\mathbf{x}) ) \\
f_2[\boldsymbol{\lambda}_2] \cdot f_1[\boldsymbol{\lambda}_1](\mathbf{x}) = G_d (y_1^{'}, y_2) \\
x_1^{'}, y_2^{'} = G_e (f_2[\boldsymbol{\lambda}_2](\mathbf{x}) ) \\
f_1[\boldsymbol{\lambda}_1] \cdot f_2[\boldsymbol{\lambda}_2](\mathbf{x}) = G_d (y_1, y_2^{'} )
\end{gather}
\end{minipage}

The overall objective is to minimize both the reconstruction error and supervenient commutativity error between $f_1$ and $f_2$.  
Encoder $G_e$ and decoder $G_d$ are neural networks and the synaptic weights are trained by loss function (equation  \ref{loss} ): 

\begin{minipage}{0.9\hsize}
\begin{gather}
\label{closs}
loss_c = \| f_1[\boldsymbol{\lambda}_1] \cdot f_2[\boldsymbol{\lambda}_2](\mathbf{x}) - f_2[\boldsymbol{\lambda}_2] \cdot f_1[\boldsymbol{\lambda}_1](\mathbf{x}) \|=
\| G_d(y_1^{'}, y_2) - G_d(y_1, y_2^{'}) \| \\
\label{loss}
loss = \| \mathbf{y} - G_d (y_1, y_2) \| + gain* loss_c, 
\end{gather}  
\end{minipage}

where $loss_c$ is a supervenient feedback error at the macro level to satisfy commutativity between $f_1$ and $f_2$ (Figure \ref{fig:outline}b).
In structural downward causation condition, we set $gain$ to 1. 
In the ablation condition, we set $gain$ to 0.   

\subsection{Method}

\paragraph{Dataset} \quad \\
The dataset consists of 26 alphabets with 12 fonts and 7 colors. The image size is 3 channels $\times$ 32 pixels $\times$ 32 pixels. The background color is black, (0,0,0) in (R, G, B).  The seven colors of alphabets employed consist of (0, 0, 1), (0, 1, 0),$\ldots$, (1, 1, 1). Two images are randomly sampled to make pairs of $(\mathbf{x}, \mathbf{y})$. To increase the variety of colors, a random value from 0.2 to 1 was multiplied to the color channels.

\paragraph{Encoder $G_e$}
Two latent vectors are encoded by two different encoders $G_{e1}$ and $G_{e2}$. 

\begin{minipage}{0.9\hsize}
\begin{gather}
x_1 = G_{e1}(\mathbf{x}) \\
x_2 = G_{e2}(\mathbf{x}) \\
y_1 = G_{e1}(\mathbf{y}) \\
y_2 = G_{e2}(\mathbf{y}) \\
\end{gather}
\end{minipage}

$G_{e1}$ and $G_{e2}$ are different neural networks with different synaptic weights but the network structures are the same.  
The input image is convolved using three Convolutional Neural Networks (CNN)
\cite{krizhevsky2012} 
 whose kernel size is 4, stride is 2 and padding is 1 without bias, and two linear layers follow. We used ReLU function for activation. The channels of the convolution layers were 128, 256 and 512. The output dimensions of the linear layers were 16 and 32. The number of dimension of the latent vectors was 32. 

\paragraph{Decoder $G_d$} \quad \\
The input was a tuple of two latent vectors. 
The network consisted successively of three linear layers without bias and three transposed convolution layers whose kernel size is 4, stride is 2, and padding is 1 without bias, and final layer is Conv1$\times$1 layer
\cite{lin2014}. 
The output channels of the linear layers were 128, 1024 and 4096.
The output channels of the transposed convolution layers were 128, 64, and 32.
We used the same network configuration using different initializations for comparison.

\paragraph{Training} \quad \\
We used RAdam
\cite{liu2020} 
for optimization. 
The learning rate was 1e-4 and batch size was 128. We used CUDA 11.4, PyTorch 1.10.0
\cite{paszke2019} 
and Nvidia RTX 
3080Ti for training. 
Training epochs were 400.
All program codes are available at https://github.com/Yoshiyuki-Ohmura/DownwardCausation.
 
\subsection{Result}
When feedback gain of the commutativity is 0, the synaptic weights inside $G_e$ and $G_p$ are affected by only input and output data $\mathbf{x}$ and $\mathbf{y}$. The commutative loss (equation\ref{closs}) increased (Figure \ref{fig:result}
A), because commutativity between $f_1$ and $f_2$ is not generally constrained without negative feedback control. 
In contrast, when feedback gain of the commutativity is 1, the synaptic weights are affected by the commutative equation between $f_1$ and $f_2$ and the feedback error converged to 0 (Figure \ref{fig:result} A). 
The feedback error between two conditions (gain=0 or gain=1) are significantly different regardless of the same initial synaptic weights inside all neural network modules and inputs (Figure \ref{fig:result} B, Mann-Whitney U = 10000, n=100, p $<$ 2.5e-4, two-tailed). 
This result indicates that the supervenient feedback error at the macro level can change the synaptic weights inside supervenient functions at the micro level and serves as a concrete working example of the concept of  structural downward causation. 

\section{Conclusions}
We propose that structural downward causation is a feedback control of the micro level synaptic weights using the macro level algebraic structural feedback error.
The neurons are systems that can modify their synaptic weights using the  supervenient macro-level feedback error. Algebraic structural error can be defined only at the macro level by the algebraic equation of the supervenient functions. 

Algebraic structure cannot be described at the micro level. This situation is similar to the contextual emergence \cite{Atmanspacher2007}. 
The example of contextual emergence is that the macro temperature cannot be described mathematically from the micro thermodynamics alone. New macro-level context is needed to describe the temperature. 
The context is similar to our superveneint feedback error at the macro level. However contextual emergence differs from our model in that, in contextual emergence, macro-level context cannot change the micro because molecules are not systems, they cannot change their mechanism through negative feedback control. 
Neurons, however, are systems that can change their synaptic weights  through negative feedback control. In our formulation, the macro level supervenient feedback error that cannot be described at the micro level can change the synaptic weights. 
Thus, contextual emergence is not related to structural downward causation.  

Structural downward causation is the causation from the whole level to its parts level in the structural hierarchy, and it is different from the causation from the higher level to the lower level in the functional hierarchy. Since the higher level and the lower level do not share the same physical entities, the functional downward causation is not self-causation. To our knowledge, the examples in Ellis' downward causation are considered to be functional downward causation.  \cite{ellis2009,ellis2019} 

Conventional models of downward causation mechanism \cite{rosas2020}  assume that the micro-mechanisms are fixed, and they analyze the state-to-state transition of the fixed micro-mechanism and negative feedback control is not addressed. 
In the ``macro beats micro'' model \cite{hoel2013, rosas2020}, they assume a map from a group of the micro states to the (emergent) macro states, which means that the macro states are assumed to be a summary or coarse-grained state of the micro states.  This means that the emergent hierarchy they assumed is not structural hierarchy but functional hierarchy. 
Furthermore, a modular structure to define structural hierarchy was not assumed in the conventional mechanism models \cite{hoel2013, rosas2020}. 
These are different from our whole--parts relationship, and such differences are essential to explain self-causation. 
Thus, a model of structural downward causation mechanism has yet to be proposed.

In whole--parts relationship, physical entities must be shared by the whole and its parts to explain self-causation. 
Neural network modules at the macro level and their component neurons at the micro level satisfy such a relation.
In our working model, the higher-level in the hierarchy is not a quantity that characterizes the state of the system like functional downward causation \cite{ellis2009} or coarse-grained state derived from the micro states (e.g., average of the micro states) \cite{rosas2020}, but supervenient functions that are composed of neurons and synapses.
Furthermore, the algebraic structural feedback error is described at the macro level independent of external macro cause. The external macro like an externally given reference value does not supervene on the micro. 
Therefore, the reference is not suitable for the structural downward causation.
In algebraic structure control, equations described by the supervenient functions are used to control the micro synaptic weights. 

The proposed formulation applies only to the brain or brain-like systems and is not generalizable to other objects because we assume that the micro-element is a feedback control system, not a molecule or a physical state. 
The system can perform negative feedback control autonomously using its observations, but the molecules cannot. 
In a dynamical system, the focus is often on the transition from state to state, and negative feedback control is not addressed because it is believed that negative feedback control requires a reference signal from outside the system.  Additionally, an external reference signal is not suitable for modeling  autonomous systems such as biological organisms and conscious brains. 
In the self-organization system, the interaction between particles at the same hierarchical level is the focus and the interaction between the different hierarchical levels in modular structure has not been addressed. 
In this paper, we proposed a new negative feedback control system using supervenient feedback error at the macro level without the reference signal from outside the system.  

Although our formulation of structural downward causation in the brain lacks supporting evidence, there is  also a lack of evidence that would refute the formulation. 
Structural downward causation has been considered as mysterious because no actual examples have been found, whereas reductive physicalism provides reasonable explanations for all phenomena. 
However our model of structural downward causation can be implemented in a computer simulation and we have confirmed that the supervenient macro-level feedback error can indeed change the synaptic weights at the micro level.
Therefore, we can accept the possibility that the brain uses a similar self-causation mechanism. 
We believe that some kind of self-causation mechanism is necessary to explain mental phenomena, including conscious experience, intentionality, intrinsic rewards, and volition. 

In this article, we present the first working model of self-causation. 
We allow for the possibility that other self-causation mechanisms exist. 
Our main claim is, that the mere existence of self-causation supports the possibility of  mental causation, because we can reject the A1 argument (cause and effect require separate entities). 
We expect that this research will lead to future understanding of higher-order laws, including the laws of conscious experience and psychology, and to the search for evidence of structural downward causation in living organisms. 
 
\bibliographystyle{ieeetr}
\bibliography{sample}

\paragraph{Author Contributions} Conceptualization, Y.O.; Writing -- Original Draft Preparation, Y.O.; Methodology, Y.O; Software, Y.O; Data analysis, Y.O; Writing -- Review \& Editing, Y.K.; Supervision, Y.K.

\paragraph{Conflict of Interest} The author declare no conflict of interest. 

\paragraph{Acknowledgement} We would like to thank Earnest Kota Carr for editing language.  

\end{document}